\documentclass[useAMS,onecolumn]{mn2e}
\usepackage{times}

\usepackage{graphicx}
\usepackage{amsmath}
\usepackage{amssymb}

\title{Outliers to the Isotropic Energy - Peak Energy Relation in GRBs}

\author[E. Nakar \& T. Piran]
  {Ehud Nakar,$^1$
  and Tsvi Piran,$^{1,2}$ \\
  $^1$Theoretical Astrophysics,
Caltech, Pasadena, CA 91125, USA\\
  $^2$Racah Institute for Physics, The Hebrew
University, Jerusalem 91904, Israel}

\begin{document}
\maketitle

\begin{abstract}

The peak energy - isotropic energy (EpEi) relation is among the most
intriguing recent discoveries concerning GRBs. It can have numerous
implications on our understanding of the emission mechanism of the
bursts and on the application of GRBs for cosmological studies.
However, this relation was verified only for a small sample of
bursts with measured redshifts. We propose here a test whether a
burst with an unknown redshift can potentially satisfy the EpEi
relation. Applying this test to a large sample of BATSE bursts we
find that a significant fraction of those bursts cannot satisfy this
relation. Our test is sensitive only to dim and hard bursts and
therefore this relation might still hold as an inequality (i.e.
there are no intrinsically bright and soft bursts). We conclude that
the observed relation seen in the sample of bursts with a known
redshift might be influenced by observational biases and from the
inability to locate and well localize hard and weak bursts that have
only a small number of photons. In particular we point out that the
threshold for detection, localization and redshift measurement is
essentially higher than the threshold for detection alone. We
predict that Swift will detect some hard and weak bursts that would
be outliers to the EpEi relation. However, we cannot quantify this
prediction.  We stress the importance of understanding the
detection-localization-redshift threshold for the coming Swift
detections.

\end{abstract}

\section{Introduction}
The detection of Gamma-ray Bursts (GRBs) afterglows enabled the
determination of the redshift for a few dozens bursts (out of
several thousands observed so far). This yielded a small sample of
bursts for which the {\it observed} properties can be translated
into {\it intrinsic} ones. This, in turn, initiated the search for
relations between various intrinsic properties. Such a relation can
have far reaching implications both on the theoretical understanding
of GRBs and on the application of GRBs as a tool.

Even before a large sample of bursts with redshift was available, it
was suggested that the intrinsic $E_p$ and $E_{iso}$ are correlated
(Lloyd et al. 2000, Lloyd-Ronning \& Ramirez-Ruiz 2002). Once more
than a dozen redshifts were measured, Amati et al. (2002) reported a
tight relation between the isotropic equivalent bolometric energy
output in $\gamma$-rays, $E_{iso}$, and the {\it intrinsic} peak
energy of the $\nu f_\nu$ spectrum, $E_p$ (hereafter we denote the
$E_p$-$E_{iso}$ relation as EpEi):
\begin{equation}
E_{iso}=A_k E_p^k \ , \label{EQ EpEi}
\end{equation}
where $k \sim 2$ and $A_k$ is a constant. This result was based on a
sample of 12 BeppoSAX bursts with known redshifts.  Ten additional
bursts detected by HETE II (Lamb et al., 2004; Atteia et al., 2004)
supported this result and extended it down to $E_{iso} \sim
10^{49}$ergs (see also Ghirlanda et al. 2004a).

Two bursts, within the current sample of bursts with a known
redshift, GRB 980425 and GRB 031203 are clear outliers to the EpEi
relation. Both are dim (low $E_{iso}$) and hard (high $E_p$). These
two bursts are usually ignored as a peculiar outliers to a confirmed
relation. Even though the EpEi relation is based on a small and
unique sample (bursts with confirmed redshift and a well observed
spectrum), and even though there are two clear outliers, this
relation initiated numerous attempts to explain it theoretically and
to use it for various applications. Therefore, testing the validity
of the EpEi relation with the largest available sample (of several
thousands BATSE bursts), is extremely important. This is the goal of
this letter.

We present here (Eq. \ref{EQ d}) a simple test whether a burst can
potentially satisfy the EpEi relation. This test can be carried out
for bursts with unknown redshift as long as we have a lower limit on
the observed peak energy, $E_{p,obs}$ and an upper limit on the
observed bolometric fluence, $F$. A burst that fails this test must
be an outlier satisfying: $E_{iso} < A_k E_p^k$. On the other hand a
burst that passes this test does not necessarily satisfy the EpEi
relation. One of the known outliers, GRB 980425, fails the test only
marginally. However its low measured redshift puts it as a clear
outlier.

First, we apply the test to a larger, but yet limited, sample of
$63$ BATSE bursts with unknown redshifts and a good spectral data
(taken from  Band et al. 1993 and Jimenez, Band, \& Piran, 2001). We
find that at least $\sim 25\%$ out of these bursts significantly
fail the test and therefore are essentially outliers to the EpEi
relation. Next, we consider the full current BATSE catalog
\footnote{http://www.batse.msfc.nasa.gov/batse/grb/catalog/current/},
for which we test all the long bursts ($T_{90}>2sec$) with a
complete fluence data in all the four energy channels. The exact
spectrum for these bursts is unknown, but we can still use the BATSE
four energy channels data to obtain a lower limit on $E_{p,obs}$ for
about half of the bursts. We find that $\sim 25\%$ of the bursts in
the BATSE sample fail the test, and must be outliers to the EpEi
relation. The large numbers of outliers that we find in the
different samples of BATSE bursts, suggest that the EpEi relation is
not a generic property of GRBs. Our results do not, however, rule
out possible correlation between $E_p$ and $E_{iso}$. We also do not
test here the recently suggested relation between $E_p$ and the
beaming-corrected energy (Ghirlanda et al. 2004a), see however Band
and Preece (2005).

In \S 2 we present the basic ideas of our analysis. We  apply the
test to the sample of BATSE bursts with a known peak energy  in \S 3
and to the whole BATSE catalog in \S 4. We discuss the implications
of this result as well as possible reasons why so few   outliers
were found in the samples of bursts with known redshifts in \S 5.

\section{Trajectories on the $E_{iso}$-$E_p$ plane\label{Trajectory}}

\begin{figure}
{\par\centering \resizebox*{0.7\columnwidth}{!}
{\includegraphics{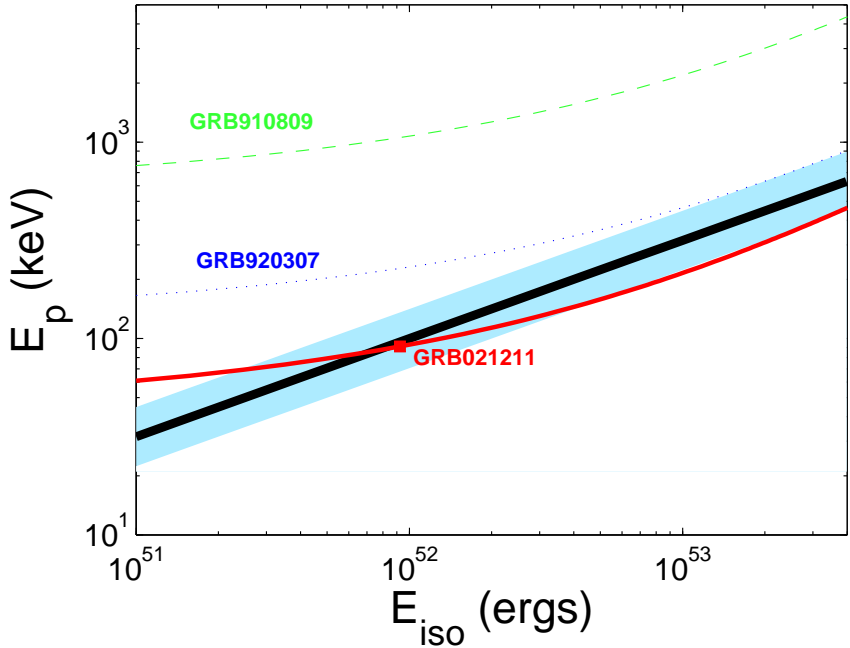}}
\par} \caption{\label{Fig_trajectory}. Trajectories of three bursts
from Band et al (1993)  and Sakamoto et al (2003) on the
$(E_{iso},E_p)$ plane. For low redshift values the trajectory is on
the left side of the figure as $E_p \rightarrow E_{p,obs}$ while
$E_{iso} \rightarrow 0$. As $z$ increases both $E_p$ and $E_{iso}$
increase (asymptotically both increase linearly with $z$) and the
trajectory moves to the upper right. The trajectory of GRB
021211(solid curve) represent a trajectory of a burst consistent
with the EpEi relation (for $k=2$, with $A_2 = 1^{+1}_{-.5} \cdot
10^{48}$ergs/keV$^2$) as it intersects the EpEi curve (gray region).
The exact position of GRB 021211 (for which the redshift is known,
z=1) on this trajectory is marked with a full square. The trajectory
of GRB 910809 (dashed curve) represent a trajectory of a burst
inconsistent with the EpEi relation. It does not intersects the EpEi
curve for any value of z. The trajectory of GRB 920307 (dotted
curve) is marginally consistent with the EpEi relation. }
\end{figure}

Consider a burst with  known bolometric fluence,  $F$, and observed
peak energy,$E_{p,obs}$, but an unknown redshift, $z$. Assuming a
$z$ value  we can evaluate the intrinsics $E_{iso}$ and $E_p$. The
trajectory of the burst on the $(E_{iso},E_p)$ plane as we vary z is
given by:
\begin{eqnarray}
 E_{iso} &=& 4 \pi D^2 \tilde r_c^2(z) (1+z) F \label{EQ Ep}\\
 E_p &=& (1+z) E_{p,obs}  \label{EQ Ei}\ ,
\end{eqnarray}
where $D\equiv c/H_0$ and $\tilde r_c(z)$ is the dimensionless
comoving distance to redshift $z$.  This trajectory represent all
the possible values of the intrinsic $E_p$ and $E_{iso}$ for  given
$E_{p,obs}$ and $F$. On these trajectories $E_p \propto E_{iso}^0$
for small $E_{iso}$ values while $E_p \propto E_{iso}$ for
asymptotically large values of $E_{iso}$. Several such trajectories
are plotted in Fig. \ref{Fig_trajectory}.

The EpEi relation (Eq. \ref{EQ EpEi}) is represented by a curve on
the $(E_{iso},E_p)$ plane. For  $k\ge 1$ (which is satisfied by any
reasonable fit to the observed data) there are values of
($F,E_{p,obs}$) for which the trajectories (on the $E_{iso},E_p$
plane) do not intersect the EpEi curve for any value of $z$. These
trajectories correspond to outliers to the EpEi relations (which is
not satisfied for any value of $z$). Put differently, one can
imagine using the EpEi relation to determine the redshift of
observed bursts. For the bursts that the trajectories do not
intersect there will be no value of z for which the EpEi relation is
satisfied (Ghirlanda et. al. 2004b). A second group of $F,E_{p,obs}$
values are these for which the trajectories intersect the EpEi line.
These bursts can potentially satisfy  the EpEi relation as there is
a possible $z$ value for which this relation can be satisfied. Fig.
\ref{Fig_trajectory} illustrates the two types of trajectories.

Substituting Eqs. \ref{EQ Ep} \& \ref{EQ Ei} in Eq. \ref{EQ EpEi} we
obtain a general condition for an intersection between a trajectory
of an observed burst and the EpEi line:
\begin{equation}
{A_k \over 4 \pi D^2} {E_{p,obs}^k \over F} = {r_c^2(z) \over
(1+z)^{k-1}} \ .
\end{equation}
The dimensionless function on the r.h.s. depends only on $z$. It
vanishes as $z$ vanish and at large values of $z$ (for $k>1$) and
hence it has some maximal value denoted $C_k$. All the bursts for
which the observables on the l.h.s. are larger than this maximal
value are outliers to the EpEi relation. We define a ratio
\begin{equation} \label{EQ d}
d_k \equiv {A_k \over 4 \pi D^2 C_k} {E_{p,obs}^k \over F} \ .
\end{equation}
\begin{itemize}
\item Bursts with $d_k<1$ can potentially satisfy the EpEi
relation. \item Bursts with $d_k>1$ cannot satisfy the EpEi
relation. For these bursts $d_k$ is a measure of the minimal
``distance" of the burst from the EpEi relation. Namely, the
observed combination $E_p^k/F$ should decrease by this factor in
order that the EpEi relation would be potentially satisfied.
\end{itemize}

\section{Bursts with a known Observed Peak Energy}

Following the observations (Amati et al., 2002; Lamb et al., 2004;
Atteia et al., 2004) we present here (and in \S 4) the results for
$k=2$ with $A_2 = 1^{+1}_{-.5} \cdot 10^{48}$ergs/keV$^2$. The error
introduced here is our estimate of the spread in the data.  All the
bursts in the sample of Atteia et al. (2004) are consistent within
$1\sigma$ with these values. Our results do not change qualitatively
for other values of $k$ and $A_k$ that yield a reasonable fit to the
data. The cosmological parameters that we consider are
$\Omega_m=0.3$, $\Omega_\Lambda=0.7$ and $h=0.7$, for which $C_2=
0.56$. For these values we obtain:
\begin{equation} \label{EQ d2}
d_2= 8 \cdot 10^{-10} {(E_{p,obs}/1 keV)^2 \over (F/(1 erg\;
cm^{-2}))}
\end{equation}

We consider a sample of BATSE bursts (from Band et al., 1993, and
Jimenez et al., 2001) with unknown redshifts for which the observed
peak energy has been determined. We consider only bursts with a high
spectral index smaller than $-2$ in order to ensure that the break
energy in the observed spectrum is indeed the peak of $\nu F_\nu$.
Our sample includes $63$ ($40$ bursts from Band et al. 1993, and 23
bursts from Jimenez et al. 2001). Using the spectral fits for these
bursts we derive their bolometric fluence (0.1-10000 keV).

\begin{figure}
{\par\centering \resizebox*{0.7\columnwidth}{!}
{\includegraphics{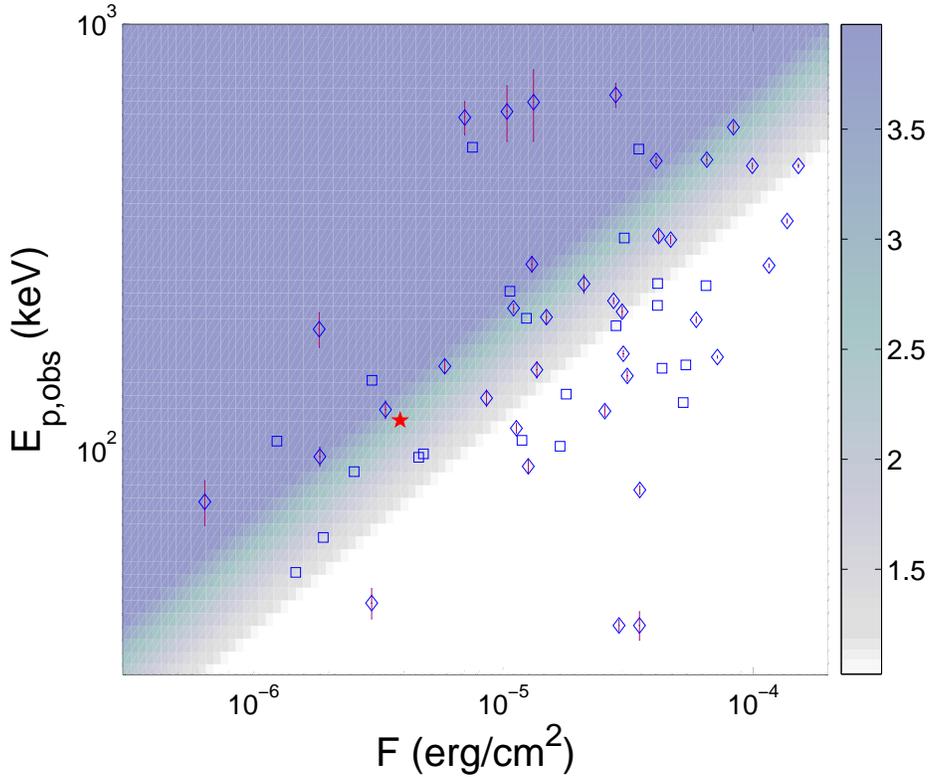}}
\par} \caption{\label{Fig_C2}. A color map of $d_2$. The region marked
in white, where $d_2<1$  corresponds to allowed solutions of the
EpEi relation. Larger values are marked with darker colors and they
correspond to the minimal ratio between $E_{iso}$ given by the EpEi
relation and $E_{iso}$ given by the $(E_p,E_{iso})$ trajectory, for
the same value of $E_p$. Also marked on this figure are values of
$F$ and $E_{pobs}$ for 39 BATSE bursts from Band et al., (1993)
(diamonds), and 22 BATSE bursts from Jimenez et al. (2001)
(squares). For 29 [15] out of these 61 bursts $d_2
> 2 [4]$. GRB 980425 (full star) has a marginal $d_2 \approx 3$.}
\end{figure}

Fig. \ref{Fig_C2} depicts a color map of $d_2$ for each burst on the
$F,E_{p,obs}$ plane. The observed values of our sample (including
error bars where available) are marked on this map. From Fig.
\ref{Fig_C2} it is evident that a significant fraction of the bursts
cannot satisfy the EpEi relation. Fig. \ref{Fig hist} depicts a
histogram of the fraction of bursts with $d_2$ larger than a given
value. We account for uncertainties in the measurement of
$E_{p,obs}$, when possible, by using an $E_{p,obs}$ value that is
smaller by $1\sigma$ than the measured value (unfortunately we can
do it only for the Band et al. 1993 sample since the uncertainties
in the measurement of $E_{p,obs}$ are not reported in Jimenez et
al., 2001). Fig. \ref{Fig hist} shows that $\approx 40\%$ of the
bursts have $d_2>2$ while $25\%$ of the bursts have $d_2>4$ (9/40
from Band et al., 1993 and 6/23 from Jimenez et al., 2001). Since
the scatter in the EpEi relation is a factor of 2 we consider,
conservatively, a burst with $d_2>4$ as an outlier. Finally, $13\%$
of the bursts are very far from the relation having $d_2>10$. We
stress that these are only lower limits. While bursts for which
$d_2<1$ can satisfy the EpEi relation, they do not necessarily do
so.

\section{BATSE Bursts}

Only a small fraction of BATSE bursts have a published $E_{pobs}$
values. Still we can obtain a lower limit of $E_{pobs} > 250$keV for
all BATSE bursts for which:
\begin{equation}
{F_{300,2000} \over F_{20,50} + F_{50,100}+F_{100,300}} > 1.25  \ ,
\label{IneqEp}
\end{equation}
where $F_{E_1,E_2}$ is the fluence between $E_1$ and $E_2$ reported
in the four BATSE windows. This lower limit holds for a Band spectra
(Band et al. 1993) over a wide range of low and high spectral
indices ($\alpha$ and $\beta$ respectively). As a test of the
validity and robustness of this criterion we apply it to the BATSE
bursts with known $E_p$ (Band et al. 1993 and Jimenez et al. 2001,
including those with $\beta>-2$ and those with known redshift). We
find that indeed all the bursts  in the sample, apart for one, that
satisfy Eq. \ref{IneqEp} have $E_{p,obs} > 250$keV (23 bursts all
together). Using this lower limit on $E_{p,obs}$ we can obtain a
lower limit on $d_2$ for a large sample of BATSE bursts, where we
take $F$ in $20-2000$ keV  energy range (the sum of all 4 channels)
as the bolometric fluence.

\begin{figure}
{\par\centering \resizebox*{0.7\columnwidth}{!}
{\includegraphics{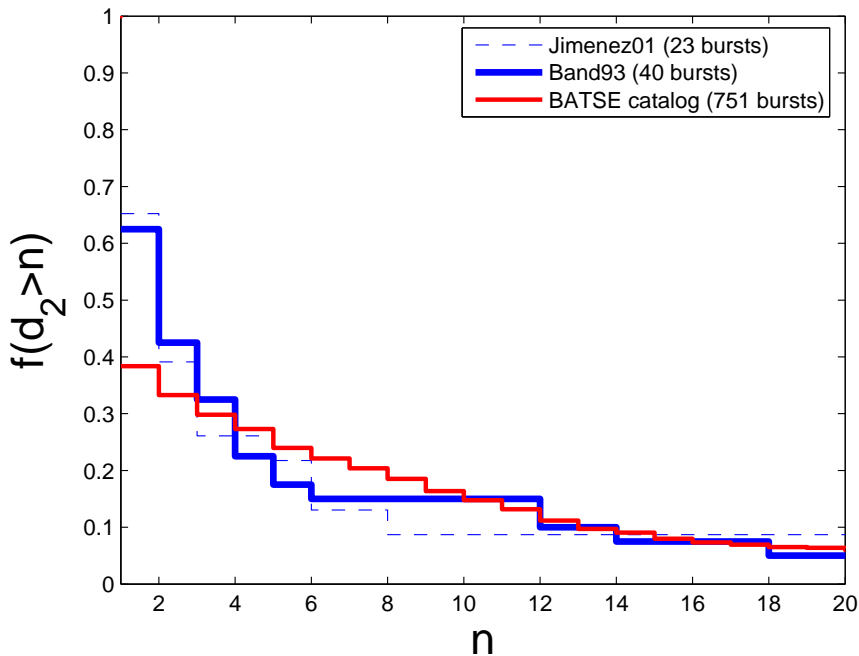}}
\par} \caption{\label{Fig hist}. A cumulative fraction of
BATSE bursts with $d_2>n$ as a function of $n$ from the samples of
Band et al 1993 (thick line), Jimenez et al. 2001 (dashed line) and
the current BATSE catalog (thin line). In the last sample (BATSE
catalog) $E_{pobs}$ was taken as larger than $250$keV for any burst
that satisfies Eq. \ref{IneqEp}.}
\end{figure}

We consider a sample of $751$ long ($T_{90}>2$sec)  bursts  from the
current BATSE catalogue. Our selection criteria were having fluence
in all four BATSE bands, having errors that are smaller than half of
the measured values in all the four bands and having a measured
$T_{90}$. $361$ of these bursts satisfy Eq. \ref{IneqEp} yielding a
lower limit on their $E_p$. Fig. \ref{Fig hist} depicts also the
fraction of long bursts out of the sample of $751$ bursts, that
satisfy Eq. \ref{IneqEp} and have $d_2>n$. We find that
approximately $35\%$ of these bursts have $d_2>2$, about $30\%$ have
$d_2>4$ and for $10\%$ this ratio is larger than 15!. While this
estimate is less robust than the previous ones (i.e. we cannot
quantify the error in the lower limit we obtain for $E_{p,obs}$) it
is clear that a significant fraction of long BATSE bursts cannot
satisfy the EpEi relation. This result has been confirmed by Band
and Preece (2005) that use a sample of 760 BATSE bursts where
$E_{p,obs}$ is known.

Finally, we have also  performed the same test for the 187 short
($T_{90}<2$sec) BATSE bursts satisfying the same criteria. These
bursts are typically harder than long ones. As they are shorter they
also have a lower overall fluence. One could expect that they won't
satisfy the EpEi inequality. We find that more than 75\% of BATSE
short bursts have $d_2>10$. Short bursts cannot satisfy the EpEi
relation! This result is similar to the one obtained by Ghirlanda et
al. (2004).

\section{Discussion}

We have presented a simple method for  testing whether a burst can
{\it potentially} satisfy the $E_p$-$E_{iso}$ relation. This method
requires only two observables, the bolometric $\gamma$-rays flux and
the peak energy. Both can be determined for every observed burst
regardless of its localization and redshift determination. We have
carried this test for several samples of BATSE bursts. We find that
$\approx 25\%$ of the BATSE bursts in these samples fail the test
and hence they are outliers to the EpEi relation. We stress that
this fraction is only a lower limit as bursts that pass the test may
still not satisfy the EpEi relation, once their redshift is known.
These results imply that the EpEi relation, in its current form, may
not a generic property of GRBs. It is present only in the small
sample of bursts with confirmed redshifts and not in the whole
sample of observed bursts.

None of the outliers we find has an isotropic energy larger than the
one predicted by the EpEi relation. Truly, our test could not find
such bursts. However, the two known outliers have lower isotropic
energy than the one predicted by the EpEi relation. Moreover,
already the BATSE data demonstrated the absence of soft and bright
bursts. The absence of such bursts is confirmed by BeppoSAX and HETE
II which would have easily detected and localized them. Thus, we
suggest that the common EpEi relation should be replaced by an EpEi
inequality:
\begin{equation}
E_{iso} \lesssim A_k E_p^k \ .
\end{equation}

The natural question that arises  is why there are so many
outliers in the BATSE data while there are only two outliers to
the EpEi relation in the current sample of bursts with confirmed
redshifts? One possibility is that there are systematic errors.
Since $d_2 \propto E_{p,obs}^2$, if for some reason $E_{p,obs}$ of
all the BATSE bursts is overestimated by a factor of $\gtrsim 2$
or if it is underestimated by the same factor for BeppoSAX and
HETE II bursts, then BATSE sample may be consistent with an EpEi
relation. The other possibility is that the difference between
BATSE data and the current sample of bursts with confirmed
redshifts results from an observational selection effect
(Lloyd-Ronning \& Ramirez-Ruiz 2002). This explanation is
supported by the fact that both outliers were not localized in the
usual manner by either BeppoSAX or HETE II whose localized bursts
compose the localized bursts sample. The first, 980425, was
detected and localized by BeppoSAX. However, if it was not for the
discovery of SN 1998bw (Galama et al., 1998) the identification of
its host galaxy and the measurement of its redshift would have
remained questionable. The second outlier, 031203 was localized by
INTEGRAL (Sazonov, Lutovinov \& Sunyaev 2004). Observational
selection affects might play a complicated roll especially since
the threshold for redshift measurement might be higher than the
threshold for detection. This is intuitively clear as the redshift
determination requires not only a detection of the prompt emission
but also a fast localization and an afterglow detection.

Our results suggest that Swift, which is expected to reduce the
threshold for detection, localization and afterglow detection,
will detect dim and hard bursts that do not satisfy the EpEi
relation. It is impossible, however, to quantify this prediction
without a clear understanding of the threshold for redshift
measurement. Moreover, this second threshold would have to be
understood in order to use the coming sample of Swift bursts with
known redshifts to study the relation between $E_p$ and $E_{iso}$,
or other intrinsic properties of the GRB population.

\vspace{0.25in}

This work was supported  by a BSF grant to the Hebrew University. We
thank Enrico Ramirez-Ruiz, Eli Waxman, Dafne Guetta, Amir Levinson
and Re'em Sari for helpful discussions. We thank the anonymous
referee for his detailed and constructive report.

\end{document}